# Single-shot structured illumination microscopy

QINNAN ZHANG,[1,2,§] EN BO,[3,§] JIAWEI CHEN,[1,§] JIAOSHENG LI,[4] HEMING JIANG,[3] XIAOXU LU,[4] LIYUN ZHONG,[4] JINDONG TIAN[1,*]

[1] *College of Physics and Optoelectronic Engineering, Shenzhen University, Shenzhen 518060, China*
[2] *School of Electrical Engineering and Intelligentization, Dongguan University of Technology, Dongguan 523808, China*
[3] *MGI, BGI-Shenzhen, Shenzhen 518083, China*
[4] *Guangdong Provincial Key Laboratory of Nanophotonic Functional Materials and Devices, South China Normal University, Guangzhou 510006, China*
[§] *These authors contributed equally to this work*
*Corresponding author: jindt@szu.edu.cn*

**Structured illumination microscopy (SIM) can double the resolution beyond the light diffraction limit, but it comes at the cost of multiple camera exposures and the heavy computation burden of multiple Fourier transforms. In this paper, we report a novel technique termed single-shot SIM, to overcome these limitations. A multi-task joint deep-learning strategy is proposed. Generative adversarive networks (GAN) are employed to generate five structured illumination images based on the single-shot structured illumination image. U-Net is employed to reconstruct the super-resolution image from these six generated images without time-consuming Fourier transform. By imaging a self-assembling DNB array, we experimentally verified that this technique could perform single-shot super-resolution reconstruction comparing favorably with conventional SIM. This single-shot SIM technique may ultimately overcome the limitations of multiple exposures and Fourier transforms and is potentially applied for high-throughput gene sequencing.**

## 1. INTRODUCTION

Imaging of self-assembling DNA nanoball (DNB) array is an important step in the high-throughput gene sequencing [1-4]. The DNA bases are labeled with four specific fluorescent dyes. And a specific fluorescence microscopy is employed to detect the fluorescent signals. The throughput of the gene sequencing technology is determined by the imaging speed and the number of DNB that can be identified by fluorescence microscopy in the field of view (FOV). Therefore, the spatial resolution and imaging speed directly affects the throughput of the gene sequencing techniques. However, the Abbe diffraction limit [5] limits the spatial resolution of the fluorescence microscope at only half the wavelength of incident light.

With the development of microscopy, super-resolution fluorescence microscopy technique provides the possibility for the breakthrough of the diffraction limit, such as stochastic optical reconstruction microscopy (STORM) [6, 7], photoactivated localization microscopy (PALM) [8, 9], structured illumination microscopy (SIM) [10, 11], stimulated emission depletion (STED) [12, 13], and other super-resolution (SR) microscopy [14–16]. Owing to its advantage of high speed, SIM stands out among these techniques. SIM can double the resolution beyond the light diffraction limit by using wide-field illumination with multi-frame different periodic patterns SIM images [17, 18]. The high spatial frequency components, which are beyond the range of optical transfer function (OTF) in the Fourier domain, can be recorded with various phases of the illumination pattern. By using a reconstruction structured illumination microscopy algorithm [19–21], the high spatial frequency components can be shifted into the true positions and assembled in the Fourier domain. Using the nonlinear effect of fluorescence, SIM can reach more times resolution [18].

To reconstruct an SR image, SIM requires the precise frequency and phase of illumination pattern to ensure the imaging quality and Fourier transform [20] or shifting-phase SIM [22] is needed to reconstruct the SR image from three images with shifting illumination patterns at each given orientation. The conventional SIM needs nine-frame images at three given orientations, which means that the sample needs to be repeatedly exposed. Although, a frequency domain SIM algorithm is developed, which requires four raw SIM images [23]. It still limits the imaging speed and has phototoxic effects. How to improve the speed of the imaging process and reduce phototoxic effects remains an open and challenging question. It has great significance for improving the throughput of the gene sequencing technology.

Recently, deep learning (DL) as a powerful machine learning technique has attracted broad attention. The DL framework can be trained to solve traditional problems in several optical fields, such as optical imaging [24-26], digital holographic reconstruction [27,28], digital holographic microscope [29], fringe pattern analysis [30], phase unwrapping [31-33] and so on [34-36]. And DL has been proven to be useful for improving SIM. The number of raw images required for SIM can be reduced to 3 frames [37, 38]. And DL enables SIM with low light levels and enhanced speed [37]. They allow for a higher SR imaging speed with reduced photobleaching [38].

Here, we propose a DL-based single-shot SIM technique, which can reconstruct an SR image using only one frame structured illumination image and multi-task joint learning method. The multi-task joint learning consists of two parts: Generative adversarial network (GAN)

[39] and deformation of U-Net (DU-Net) [40]. The GANs are used to generate the structured illumination images, which are needed to reconstruct the SR image. The DU-Net is used to reconstruct the SR image from these generated images without time-consuming Fourier transforms. In this paper, the proposed technique is used for imaging a 2D self-assembling DNB array [1], which is used for gene sequencing [1-4]. The new generation sequencing technology combines highly efficient imaging on ordered arrays with inexpensive ligation-based chemistry, which can greatly reduce the time and reagent cost with sufficient accuracy. The array fluorescent signal can be read by a microscope and a digital camera. We test the proposed technique on the self-assembling DNB array. The experimental results demonstrate the feasibility of the proposed technique and show the significant applications value of the proposed technique in improving the throughput and measurement speed of a single DNB array sample in the gene sequencing process.

## 2. METHODS

### A. DL-based single-shot SIM

We combine five GANs and one DU-Net to carry out multi-task joint learning. First, the GANs are used to generate the SIM raw data needed to reconstruct the SR image. According to the characteristics of the self-assembling DNB array, only six SIM raw images with three phases in two perpendicular directions are needed to reconstruct the SR image of the DNB array. Therefore, five GANs are needed to generate other five specific SIM raw images. Next, six SIM raw images are input the DU-Net to generate the final SR image. The DU-Net has six encoder channels and one decoder channel, which is better adapt to the technique in this case. We set up six separate encoder channels for each input raw image, which is used to extract the feature information in each image to the maximum extent. Finally, all feature information of the six raw images is integrated into the decoder to map the final SR image. The schematic of the proposed technique is shown in Fig. 1. The size of an individual DNB is 220 nm, the central wavelength of the fluorescence signal is 550 nm, and the distance between the DNB is 480 nm. In our imaging system, the numerical aperture of the objective is 0.8. And the theoretical resolution of the imaging system is 420 nm.

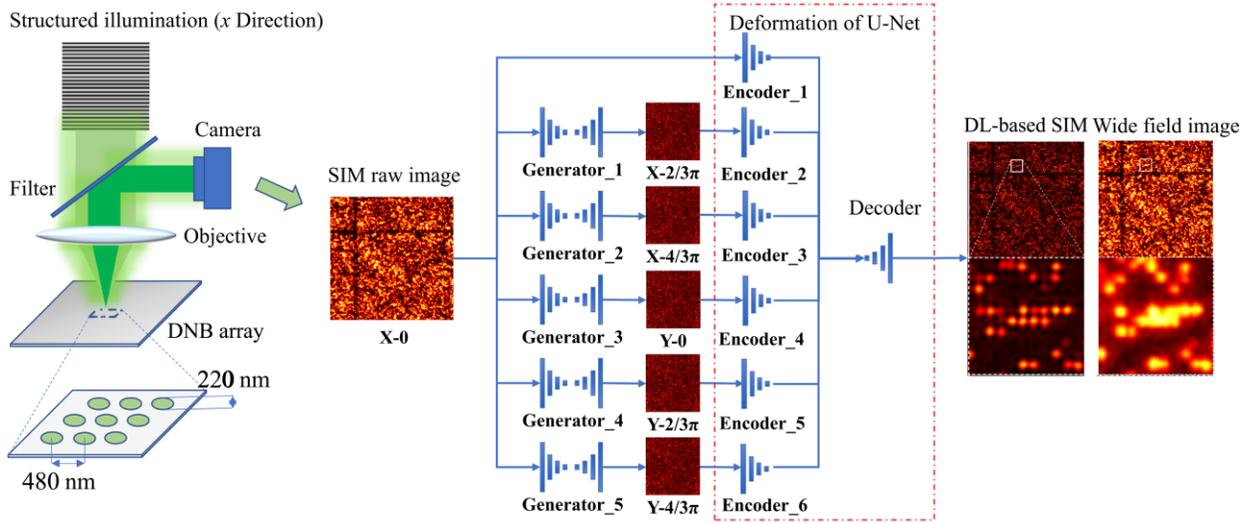

Fig. 1. The schematic of DL-based single-shot structured illumination microscopy (SIM) for imaging the self-assembling DNB array.

### B. Multi-task joint learning

We integrate five GANs altogether, which have five generator networks and five discriminator networks. The training will optimize the weight parameters sequentially to achieve simultaneous training. During the training, the SIM raw image with the initial phase of zero is put into the generator network as input. The generator network is a conventional U-Net [40], which adopts a convolution kernel of size 3×3 in each layer to extract depth feature information with the step of 2 pixels, and adopts the convolution kernel with the size of 5×5 to check the results of the under-sampling for deconvolution with the step of 2 pixels. The discriminator network is a convolutional neural network (CNN), which classifies the input image and finally outputs a scalar indicating whether the input image is fake or real. The loss function is

$$Loss_{GAN} = E_{x\sim p}\left[\log D(X)\right] + E_{x\sim p}\left[\log\left(1 - D(G(Z))\right)\right], \quad (1)$$

where $D(X)$ represents the discriminator's judgment of label data ($X$), and $G(Z)$ represents the output data of the generator. The adversarial strategy is used to training five GANs [35]. The generator network and the discriminator network fight against each other and continue to learn so that the image generated by the generator network cannot be recognized by the discriminating network as real or fake, and the training can be considered complete at that time. After training, the generate_1 and generate_2 are used to shift the phase of the SIM raw images with illuminating direction X, generate_3 is used to change the illuminating direction from X to Y, generate_4 and generate_5 are used to generate the phase-shifting SIM raw images with illuminating direction Y. After that, one input SIM raw image and five generate SIM raw images are put into the deformation of U-Net. The architecture of the network is shown in Fig. 2. Six encoders are used to extract the features of each raw image. And the features of each layer are combined for the decoding of the SR images.

The Root-Mean-Square Error (RMSE) between the generated super-resolution image and the label SR image is used as the Loss of the DU-Net.

$$Loss_{DU\text{-}Net} = \sum_{p\times q} \frac{\left[G(x,y) - X(x,y)\right]^2}{p\times q}, \quad (2)$$

where $G(x,y)$ represents the output data of DU-Net, $X(x,y)$ represents the label super-resolution image, p and q represent the size of the input and label images. x and y are the pixel coordinates of the images.

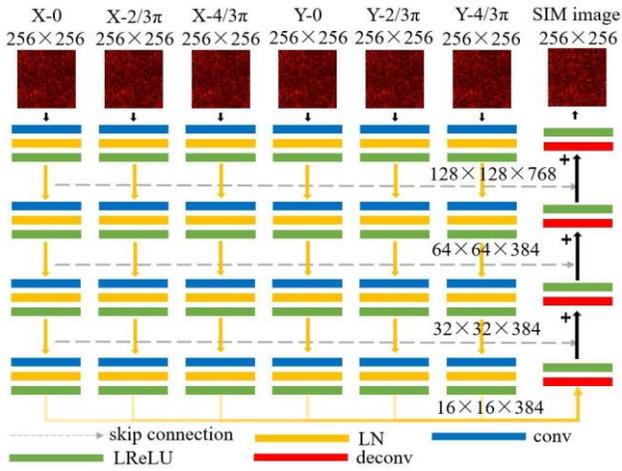

Fig. 2. The architecture of the deformation of U-Net (DU-Net).

### C. Generation of the training data

A self-built structured illumination microscope is used to generate the training data. Six SIM raw images are recorded for the reconstruction of an SR image in the *x* and *y* directions. Three raw images with 0, 2/3π, and 4/3π phase shifting are recorded in each direction. The SIM raw image with 0 phase shifting at *x*-direction inputs five GANs in sequence as the input data. The remaining SIM raw images are used as the label for the training of GANs. Five GANs have generated the raw images with 2/3π and 4/3π phase shifting at *x*-direction and the raw images with 0, 2/3π, and 4/3π phase shifting at *y*-direction. And then, to training the DU-Net, the recorded SIM raw images and the corresponding SR reconstructed image are used as the input data and the label. The OpenSIM algorithm [20] is used to reconstruct the super-resolution image from six recorded SIM raw images. And we record 10000 groups of images with 80% for training 20% for testing. All training was performed on the cloud server Intel(R) Xeon(R) CPU E5-2678, with 32 GB of RAM and NVIDIA TITAN Xp.

### D. Gene sequencing chip preparation

The preparation process of gene sequencing chip is mainly divided into three steps:

**Step1: DNBs preparation**

The standard reference *E. coli* or NA12878 DNA samples were used to prepare DNBs. The number of copies of the DNA template was increased based on rolling circle amplification technology via the DNBSEQ DNB Make Reagent Kit (MGI Technologies, Shenzhen, China; Product number 1000027737). As the regular array density of the chip used in this experiment is relatively high, and the size of DNB needs to be controlled. During the DNA rolling circle replication process, the time can be reduced to 10 minutes to obtain a DNB of a suitable size and normal concentration.

**Step2: DNB loading**

DNBs obtained in Step1 need to be loaded onto a high-density regular array chip. Add loading buffer to the DNB mixture, pump it into the chip with the help of a mini-loader in a volume of 30μl/lane, and leave it at room temperature for 30 minutes so that DNBs can settle on the chip well and get a chip regularly covered with DNBs.

**Step3: DNB consolidation and hybridization with fluorescently labeled dNTP**

After DNBs are loaded on the chip, use the gas-liquid instrument system to pump different reagents into the chip so that DNBs can be loaded on the chip more tightly and firmly (DNB consolidation), and then pump the sequencing prime and the sequencing reagents sequentially. The sequencing prime and the DNB adapter complementarily hybridize. The sequencing prime is combined with the fluorescently labeled dNTP in the sequencing reagent under the catalysis of DNA polymerase. DNBs on the chip emit fluorescence when it is excited by the light source, and the fluorescence is collected by the imaging system for gene sequencing.

The study was approved by the Institutional Review Board IRB-BGI (IRB-202104090286).

## 3. EXPERIMENTAL RESULTS

We validate the proposed technique on experimental data. The fluorescence image of labeled base A is shown in Fig. 3. The central wavelength is 550 nm. And the theoretical resolution of the imaging system is ~420 nm. Wide-field image, reconstructed SR image from six recorded raw images with OpenSIM algorithm (OpenSIM 6), reconstructed SR image from one recorded raw image and five generated raw images with OpenSIM algorithm (OpenSIM 1) and the reconstructed SR image from one recorded raw image and five generated raw images with DU-Net are shown as in Fig. 3(a) to 3(d). Comparing with wide-field images, the plots of the white line in the partially enlarged detail are shown as in Fig. 3(i). The full width at half maximum (FWHM) of a single signal point is 550 nm in a wide-field image. The FWHMs of SR images respectively are 339 nm, 332 nm, and 335 nm. It proved that the proposed technique has effectively improved the imaging resolution comparable to traditional techniques, which require multiple SIM raw images. As shown in Fig. 3(j), a different signal point can be identified, which is indistinguishable in a wide-field image. And the corresponding frequency domain images are shown as in Fig. 3(e) to 3(h). It proved that the proposed technique can effectively improve the high-frequency information of the image and can realize the reconstruction of an SR image by recording only one SIM raw image.

And then, decorrelation analysis is used for image resolution estimation at the frequency domain [22, 41]. The algorithm requires only a non-saturated, bandwidth-limited signal with adequate spatial sampling. The algorithm uses only linear operations and enables the real-time objective assessment of image resolution. First, the Fourier transform of the image is computed after standard edge apodization to suppress high-frequency artifacts. The Fourier transform of the input image is normalized. And then the input image and the normalized image are cross-correlated in Fourier space using Pearson correlation and condensed to a single value between 0 and 1. Second, theoperation is repeated, but the normalized Fourier transform is filtered additionally by a binary circular mask of radius *r* [0,1] expressed in normalized frequencies. the decorrelation function can be expressed as follow:

$$d(r) = \frac{\int \text{Re}\{I(k)I_n^*(k)M(k;r)\}dk_x dk_y}{\sqrt{\int |I(k)|^2 dk_x dk_y \int |I_n(k)M(k;r)|^2 dk_x dk_y}}, \quad (3)$$

where $k = [k_x, k_y]$ denotes Fourier space coordinates, $I(k)$ is the Fourier transform of the input image, $I_n(k)$ is the normalized Fourier transform image, $M(k; r)$ is the binary mask of radius *r*. The core idea of the technique is that by normalizing the Fourier transform of the input image and balancing the signal and noise contributions. The information of the object structure is preserved in the phase. The highest frequency ($kcmax$) is estimated from the local maxima of the decorrelation functions. By using the decorrelation function, we can find the cut-off frequency and enabling parameter-free image resolution estimation. And the resolution can be calculated as the following equation, Resolution=(2×*pixelsize*)/*kcmax* [41]. The *pixelsize* of microscopic image is 219.5 nm. And the size of the image has been doubled in the frequency domain. The expanded area of the frequency domain image is zeroed. Therefore, the *kcmax* of images is normalized to [0, 2]. The decorrelation functions of Fig. 3(a) to 3(d) are calculated wide-field image and SR image is calculated, as shown in Fig. 4. And the corresponding resolution is shown in Table 1.

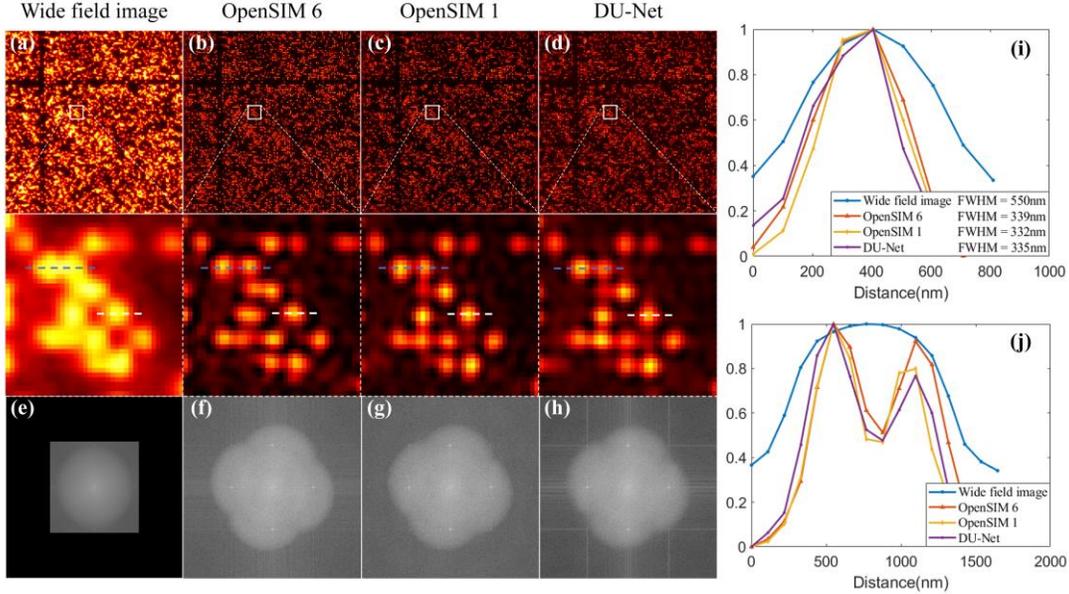

Fig. 3. Experimental comparison of base A imaging. (a) wide-field image, (b) reconstructed SR image from six recorded raw images with OpenSIM algorithm (OpenSIM 6), (c) the reconstructed SR image from one recorded raw image and five generated raw images with OpenSIM algorithm (OpenSIM 1), (d) the reconstructed SR image from one recorded raw image and five generated raw images with DU-Net, the corresponding frequency domain images (e) to (h), (i) the plots of the white line in the partial enlarged detail, and (j) the plots of the blue line in the partial enlarged detail.

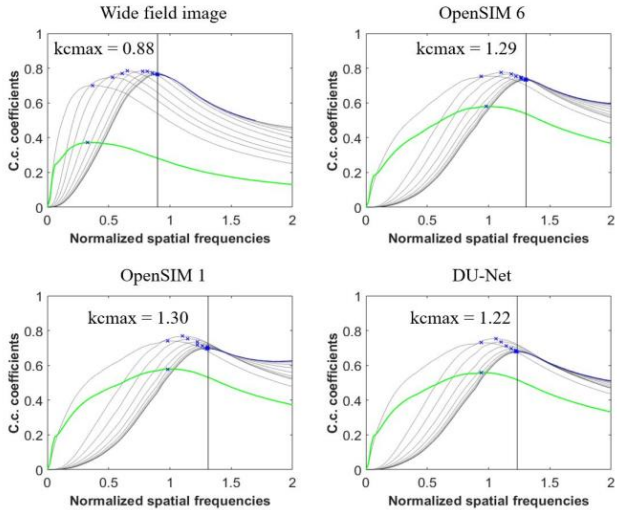

Fig. 4. Decorrelation analysis results of the (a) wide-field image, (b) reconstructed SR image (OpenSIM 6), (c) reconstructed SR image (OpenSIM 1), (d) the reconstructed SR image (DU-Net), green curves are decorrelation functions before high-pass filtering; gray curves are all high-pass filtered decorrelation functions; blue crosses are all local maxima calculated by decorrelation analysis; bold vertical lines indicating the local maximum of highest frequency *kcmax* (normalized).

As shown in the results of the decorrelation function, the resolution and FWHM show good consistency. And the resolution of the SR image, which is reconstructed with the proposed technique and one recorded raw image, is slightly reduced. However, the resolution of SR imaging is still up to 359 nm. Besides, we found that the single-shot SIM can be realized with 338 nm resolution by using OpenSIM 1 method, which reconstructed SR image from one recorded raw image and five generated raw images with OpenSIM algorithm. Comparing OpenSIM 1 method and the proposed technique, DU-Net shows extraordinary computational speed in image reconstruction process. In our case, the calculation time of DU-Net is 30 ms, and the OpenSIM algorithm needs 1 s to complete the image reconstruction. DU-Net will greatly save the time required for SR image reconstruction.

**Table 1. The highest frequency and the resolution of the wide-field image and reconstructed SR images**

|  | Wide-field image | OpenSIM 6 | OpenSIM 1 | DU-Net |
|---|---|---|---|---|
| ***kcmax*** | 0.88 | 1.29 | 1.30 | 1.22 |
| **Resolution** | 498 nm | 340 nm | 338 nm | 359 nm |

## 4. DISCUSSION AND CONCLUSIONS

In this paper, we propose a DL-based single-shot SIM multi-task joint learning which is used to reconstruct the super-resolution image from only one frame structured illumination image. GAN and DU-Net are combined to carry out multi-task joint learning. The GANs are used to generate multiple structured illumination images by input a SIM raw image. The super-resolution image can be reconstructed from these generated images without time-consuming Fourier transforms by using the DU-Net. Experimental results proved the feasibility of the proposed technique. The decorrelation function is used to analyze the resolution of the reconstructed SR images, and the resolution and FWHM show good consistency. Besides, we found that the single-shot SIM can be realized with a high resolution by combining the traditional SIM image reconstruction algorithm and our technique. It can improve the speed and device complexity of SIM at weakens phototoxicity in the imaging process. And the DU-Net can greatly save the time required for SR image reconstruction. This demonstrates that the proposed single-shot SIM technology is potentially applied for high-throughput gene sequencing.

**Funding sources and acknowledgments.**
We thank Xiaofang Wei for the gene sequencing chip (DNB array) preparation. This work was sponsored by the National Natural Science Foundation of China (NSFC) (62075140, 61805086, 61727814, 61875059, 61771138), Science, Technology and Innovation Commission of Shenzhen Municipality (202010261650007001),

**Disclosures.** The authors declare no conflicts of interest.